\documentclass{article}
\usepackage{graphicx}
\usepackage{amsmath}

\input{tcilatex}

\begin{document}

\noindent {\huge An alternative to the variation of the fine structure
constant}

{\huge \bigskip }

\noindent {\large M. E. DE SOUZA}

\noindent \noindent Departamento de F\'{i}sica, \noindent Universidade
Federal de Sergipe, Campus Universit\'{a}rio, 49100-000 S\~{a}o
Crist\'{o}v\~{a}o, Sergipe, Brazil - \noindent e-mail: mdesouza@ufs.br

\bigskip

\noindent {\large Considering that quasars evolve into normal galaxies the
average expansion velocity of matter in quasars (or galaxies) can be
estimated approximately. Such velocity makes us to see the light of quasars
slightly Doppler shifted which explains the misleading variation of the fine
structure constant.}

\textbf{\bigskip }

\noindent There have been reports$^{1-4}$ in the literature proposing a time
variation of the fine structure constant. They are based on observations of
absorption lines of gas clouds seen against background quasars. Thus the
clouds are quasar absorbers. This supposed time variation of $\alpha $
motivated the recent and very important article of Davies$^{5}$ which
proposes that the velocity of light may vary with time. This would cause
profound changes in the foundatins of physics.

\bigskip

\noindent This work presents an alternative explanation for the phenomenon
observed in the gas clouds which maintains $\alpha $ as a fundamental
constant of nature, that is, independent of time. It is based on a proposal
for the evolution of galaxies from quasars. \noindent The remarkable work of
M\'{a}rquez et al.$^{6}\;$has shown that very high redshift elliptical
galaxies appear to harbor quasars. They have also shown that such galaxies
are very small and have diameters smaller than 1kpc. All the studied objects
(about 15 quasars) have extended structures of ionized gas around them (this
fact had already been presented by other researchers). They have found other
galaxies in the fields of the studied objects only a few kpc away from them.
Some of the quasars present asymmetric radio sources with collimated
one-sided jets of extended ionized gas. This means that galaxies are born as
quasars that become galaxies by means of the shedding of matter (ionized
gas) from their cores. The same kind of phenomenon has been observed in
galaxies. Very recent data$^{7}\;$of NGC 6240, which is considered a typical
protogalaxy, show that ``approximately 70\% of the total radio power at 20cm
originates from the nuclear region ($\leq $1.5kpc), of which half is emitted
by two unresolved (R$\leq $30pc) cores and half by a diffuse component.
Nearly all of the other 30\% of the total \noindent radio power comes from
an arm-like region extending westward from the nuclear region''. A very
important property of many quasars is their brightness which can vary from
night to night. This flickering may have its origin in the outward motion of
large quantities of matter from their cores. This brightness variability is
also present in Seyfert galaxies which are powerful sources of infrared
radiation. Many of them are also strong radio emitters. For example, over a
period of a few months, the nucleus of the Seyfert galaxy M77(or NGC1068)
switches on and off a power output equivalent to the total luminosity of our
galaxy$^{8}$. It is also worth noting that the nuclei of Seyfert galaxies
are very bright and have a general starlike appearance. Researchers have
found that some Seyfert galaxies exhibit explosive phenomena$^{8}$. For
example, M77 and NGC4151 expel huge amounts of gas from their nuclei. The
spectra of both galaxies show strong emission lines, just as quasars$%
^{\prime }$. Shaver et al.$^{9}\;$ have found that there is almost no quasar
for $z<0.5$. This clearly shows that quasars evolve into galaxies.

\noindent This ejection of matter is quite common in active galaxies also.
Let us mention just some of them. NGC 2992 presents a jet-like structure and
a circum-nuclear ring$^{10}$. Falcke and Biermann$^{11}\;$ report that there
is a large scale emission-like jet going outward from the core of NGC 4258
with a mass of about $4{\times }10^{35}$kg and with a kinetic power of
approximately $10^{42}$ergs/s and expansion velocity of 2000km/s. This is of
the same order of supernovae velocities. It is well known that BL Lacertae
objects are powerful sources of radio waves and infrared radiation. They
share with quasars the fact of exhibiting a starlike appearance and of
showing short-term brightness fluctuations. As some quasars do, they also
have a nebulosity around the bright nucleus. Researchers$^{12}\;$ have
managed to obtain the spectrum of their nebulosity. \textit{The spectrum of
the nebulosity is strikingly similar to the spectrum of an elliptical galaxy}%
(M32$^{\prime }$s spectrum, in this case). In terms of the evolution above
described they are simply an evolutionary stage of a quasar towards becoming
a more quiet galaxy. \noindent Radio galaxies share with BL Lacertae objects
many of the properties of quasars. As Heckman et al.$^{12}\;$ have shown, in
the middle and far infrared (MFIR) quasars are more powerful sources of MFIR
radiation than radio galaxies. Also, there have been investigations showing
that the emission from the narrow-line region(NLR) in radio-loud quasars is
stronger than in radio galaxies of the same radio power$^{13,14,15}$.
Goodrich and Cohen$^{16}\;$ have studied the polarization in the broad-line
radio galaxy 3C 109. After the intervening dust is taken into account the
absolute V-magnitude of this galaxy becomes $-26.6\;$ or brighter, which
puts it in the quasar luminosity range. The investigators suggest that
``many radio galaxies may be quasars with their jets pointed away from our
direct line of sight''. It has also been established that radio galaxies are
found at intermediate or high redshifts and that they are clearly related to
galactic evolution because as the redshift increases cluster galaxies become
bluer on average, and contain more young stars in their nuclei. This is also
valid for radio galaxies: the higher the redshift, the higher their
activity. All these data show that a radio galaxy is just an evolutionary
stage of an active galaxy towards becoming a quiet galaxy, i.e., it is just
a stage of the slow transformation by means of an overall expansion of a
quasar into a normal galaxy.

\noindent In the light of the above considerations the nuclei of old spirals
must exhibit a moderate activity. This is actually the case. The activity
must be inversely proportional to the galaxy$^{\prime }$s age, i.e., it must
be a function of luminosity. The bluer they are, the more active their
nuclei must be. As discussed above there must also exist a relation between
this activity and the size of the nucleus(as compared to the disk) in spiral
galaxies. Our galaxy has a mild activity at its center. Most of the activity
is concentrated in a region called Sagittarius A, which includes the
galactic center. It emits synchroton and infrared radiations. Despite its
large energy output Sagittarius A is quite small, being only about 40 light
years in diameter. Besides Sagittarius A our galaxy exposes other evidences
showing that in the past it was a much more compact object: a) Close to the
center, \textit{on opposite sides of it}, there are two enormous expanding
arms of hydrogen going away from the center at speeds of 53km/s and 153km/s;
b) Even closer to the center there is the ring called Sagittarius B2 which
is expanding at a speed of 110km/s$^{(8,17)}$. It is worth noting that the
speeds are low(as compared to the velocities of relativistic electrons from
possible black holes). This phenomenon is not restricted to our galaxy.
Recent high-resolution molecular-line observations of external galaxies have
revealed that galactic nuclei are often associated with similar expanding
rings$^{18}$.

\noindent This paper will not deal with the mechanism of the expulsion of
matter from the center of quasars and will be restricted to showing what may
be happening in the case of the clouds absorption lines. It is well known
that the size of a quasar is smaller than 1kpc and that a typical galaxy has
a radius of about 50.000 light-years $\approx 5\times 10^{20}m$ and, as the
age of the Universe is about $10^{17}s$, the average expansion velocity of
matter in galaxies since the time when they were quasars is of the order of\ 
\begin{equation}
<V_{e}>\,\approx 5\times 10^{3}m/s.
\end{equation}
This figure is just a rough estimate. For very high redshift quasars the
expansion speed was at least an order of magnitude higher as it is inferred
from the data on shedding of matter from the cores of quasars and galaxies.
Therefore, the light of quasars should be Doppler shifted to higher
frequencies for the light that is coming from points close to the line of
sight. This means that the quasar light that reaches a gas cloud is slightly
Doppler shifted due to the expansion of matter with respect to a rest-frame
sitting in the center of the quasar. The shift is given by 
\begin{equation}
\frac{\Delta \nu }{\nu _{0}}=-\frac{<V_{e}>}{c}\approx -1.67\times 10^{-5}
\end{equation}
which is of the same order of magnitude of the shifts attributed to
variation of the fine structure constant around $z=1$ in the work of Webb et
al.$^{4}$. Take notice that if we attribute the above result to a variation
of the fine structure function we obtain 
\begin{equation}
\frac{\Delta \alpha }{\alpha }=\frac{1}{2}\frac{\Delta \nu }{\nu _{0}}%
=-0.835\times 10^{-5}
\end{equation}
which is a figure of the order of magnitude of that found in Webb et al.$%
^{4} $. \noindent Thus, comparing results from quasars at two different
redshifts $z_{1}$ and $z_{2}$ (with $z_{1}>$ $z_{2}$), since more active
quasars have higher redshifts, it is expected $<V_{e}>(z=z_{1})$ to be
larger than $<V_{e}>(z=z_{2})$ and, of course, $|\Delta \nu (z_{1})|>|\Delta
\nu (z_{2})|$. For very high redshift quasars $\frac{\Delta \nu }{\nu _{0}}$
may even reach $10^{-4}$ or $10^{-3}$ which is in line with the upper limit
of Cowie et al.$^{19}$ according to whom $|\frac{\Delta \alpha }{\alpha }%
|<3.5\times 10^{-4}$ for $z\sim 3$.

\noindent Therefore, the observations of the variation of the fine structure
constant may reveal, actually, how quasars evolve towards becoming quiet,
normal galaxies. And thus, the foundations of physics continue being solid
as ever!

\noindent

{\large References}

1. Marciano, W. \textit{Phys. Rev. Lett.} 52, 489 (1984)

2. Barrow, J. D. \textit{Phys. Rev. D \ 35}, 1805 (1987)

3. Damour, T. and Polyakov, A. M. \textit{Nucl. Phys. B423}, 532 (1994)

4. Webb, J. K. \textit{et al. Phys. Rev. Lett. 82}, 884 (1999)

5. Davies, P. C. W. \textit{et al. Nature 418}, 603 (2002)

6. M\'{a}rquez, I. \textit{et al. astro-ph/9810012.}

7. Colbert, E. J. M. \textit{et al. }in \textit{The Radio Emission from the
Ultra-Luminous Far-Infrared Galaxy NGC 6240},

\ \ \ \ astro-ph/9405046.

8. Kaufmann,III, W. J. in \textit{Galaxies and Quasars }(W.H.Freeman and
Company, San Francisco, 1979).

9. Shaver, P. A. \textit{et al. } \textit{astro-ph9801211}.

10. Chapman, S. C. \textit{et al. astro-ph/9810250.}

11. H. Falcke, H., Biermann, P. L.\textit{\ astro-ph/9810226}.

12. \noindent Heckman, T. M.\textit{\ et al. } \textit{Ap.J.}, 391, 39
(1992).

13. Baum, S, Heckman, T.\ M. \textit{Astrophys. J} \ 336, 702 (1989).

14. Jackson, N, Browne I. \textit{Nature}, 343, 43 (1990).

15. Lawrence, A. \textit{Mon. Not. R. Astr. Soc.}, 1992.

16. Goodrich, R. W., Cohen, M. H. \textit{Astrophys. J.} \textit{391}, 623
(1992).

17. Sofue, Y. \textit{Astro. Lett. Comm.} 28, 1 (1990).

18. \noindent Nakai, N. \textit{et al.} \textit{Pub. Astr. Soc. Japan} 39,
685 (1987).

19. Cowie, L. L., Songalia, A. \textit{Astrophys. J. 453}, 596 (1995)

\end{document}